# Vers un environnement multi personnalités pour la configuration et le déploiement d'applications à base de composants logiciels[1]

**Areski Flissi et Philippe Merle**

*Projet Jacquard INRIA Futurs*
*Laboratoire d'Informatique Fondamentale de Lille (LIFL)*
*UMR CNRS 8022 / Université des Sciences et Technologies de Lille (USTL)*
*59655 Villeneuve d'Ascq Cedex, France*
*Areski.Flissi@lifl.fr, Philippe.Merle@inria.fr*

RÉSUMÉ. *La multiplication des langages de description d'architectures, des modèles et des plates-formes de composants pose un sérieux dilemme aux architectes d'applications à base de composants logiciels. D'un côté, ils doivent choisir un langage pour exprimer des configurations concrètes qui seront déployées automatiquement sur des plates-formes d'exécution. D'un autre côté, ils désirent capitaliser leurs architectures logicielles indépendamment des langages de description et des plates-formes d'exécution. Pour résoudre ce problème, nous proposons un environnement multi personnalités pour la configuration et le déploiement d'applications à base de composants logiciels. Celui-ci est constitué d'un noyau central capturant un modèle canonique de configuration et de déploiement et d'un ensemble de personnalités adaptées aux langages et aux plates-formes. Cet article décrit l'architecture de cet environnement et discute des personnalités pour les modèles de composants CORBA et Fractal.*

ABSTRACT. *The multiplication of architecture description languages, component models and platforms implies a serious dilemma for component based software architects. On the one hand, they have to choose a language to describe concrete configurations which will be automatically deployed on execution platforms. On the other hand, they wish to capitalize their software architectures independently of any description languages or platforms. To solve this problem, we propose a multi personalities environment for the configuration and the deployment of component based applications. This environment is composed of a core capturing a canonical model of configuration and deployment, and a set of personalities tailored to languages and platforms. This paper details the architecture of such an environment and describes the personalities for the CORBA and Fractal component models.*

MOTS-CLÉS : *déploiement, configuration, composant, CORBA, Fractal.*

**1. Introduction**

Ces dernières années ont vu émerger une utilisation généralisée de l'approche orientée composant pour la construction d'applications réparties [SZY 02]. Cette approche tend à couvrir globalement l'ensemble des processus intervenants dans la production de logiciels notamment les aspects de conception, de développement, de conditionnement, de configuration, de déploiement et d'exécution. Ceci se matérialise à travers l'apparition de différents modèles et plates-formes intergiciels, soit industriels tels que les Enterprise Java Beans (EJB) [SUN 01] de SUN Microsystems, le modèle de composants CORBA (CCM) [OMG 02, WAN 01] de l'Object Management Group (OMG), le canevas .NET [PLA 02] de Microsoft, soit académiques comme le modèle Fractal [BRU 04] du consortium ObjectWeb. Chacun des modèles offre un langage dédié pour décrire la configuration et le déploiement d'applications réparties. De nombreux autres langages de description d'architectures ont été proposés dans la littérature [MED 00]. Ils offrent de nombreux avantages aux architectes parmi lesquels la capacité de décrire et de réutiliser des configurations et l'automatisation du déploiement des composants applicatifs. Cette dernière est mise en œuvre par les plates-formes d'exécution.

Toutefois la multiplication des langages de description d'architectures, des modèles et des plates-formes de composants pose un sérieux dilemme aux architectes. Ils doivent en effet choisir un langage spécifique à un modèle de composants afin d'exprimer des configurations concrètes qui seront déployées sur une plate-forme technologique supportant le modèle choisi. Mais, ils désirent également capitaliser leurs architectures logicielles indépendamment des langages de description et des plates-formes d'exécution. Ainsi l'idée principale qui se dégage est la suivante : comment permettre aux architectes de capitaliser leurs configurations (exprimées dans le langage de description de leur choix) et rendre la projection vers les plates-formes d'exécution indépendante des technologies ?

L'approche *Model Driven Architecture* (MDA) [MIL 03] préconisée par l'OMG répond en partie à ce problème grâce à la définition systématique de modèles abstraits indépendants des plates-formes (*Platform Independent Model* - PIM). Par exemple, une configuration peut être modélisée via des diagrammes d'instances et de déploiement d'UML 2.0 [OMG 03b]. La spécification « *Deployment and Configuration of Distributed Component-Based Applications* » (D&C) [OMG 03a] de l'OMG fournit également un PIM pour exprimer des assemblages de composants et leur processus de déploiement indépendamment de toute technologie orientée composant. Ce modèle est ensuite projeté grâce à des transformations de modèles vers des plates-formes technologiques cibles (*Platform Specific Model* - PSM). Malheureusement, à l'heure actuelle, UML 2.0 ou D&C sont des spécifications récentes et par conséquent n'ont pas encore de reconnaissance, du fait d'un manque d'outillage et de pratique généralisée.

Pour résoudre le dilemme des architectes, nous proposons un environnement multi personnalités, constitué d'un noyau central capturant un modèle canonique de configuration et de déploiement et d'un ensemble de personnalités adaptées aux langages et aux plates-formes. Notre approche s'inscrit dans la démarche MDA car le modèle canonique peut être assimilé à un PIM et les personnalités langages et plates-formes sont respectivement des transformations vers et depuis ce modèle.

La suite de l'article est organisée de la manière suivante. La section 2 caractérise les éléments fondamentaux des plates-formes pour composants logiciels, illustrés sur les modèles CORBA et Fractal. La section 3 expose les motivations, l'architecture et les bénéfices de notre environnement multi personnalités pour la configuration et le déploiement d'applications à base de composants logiciels. La section 4 se concentre plus spécifiquement sur le modèle de déploiement (indépendant des plates-formes) de cet environnement et ses personnalisations pour les modèles CORBA et Fractal. Enfin, la section 5 conclut sur l'intérêt de notre approche et identifie quelques perspectives de travail.

## 2. Caractérisation d'une plate-forme pour composants logiciels

Quel que soit le modèle de composants, décrire la configuration d'une application puis la déployer sur une plate-forme d'exécution nécessite de disposer d'un certain nombre d'éléments fondamentaux, comme illustré sur la figure 1.

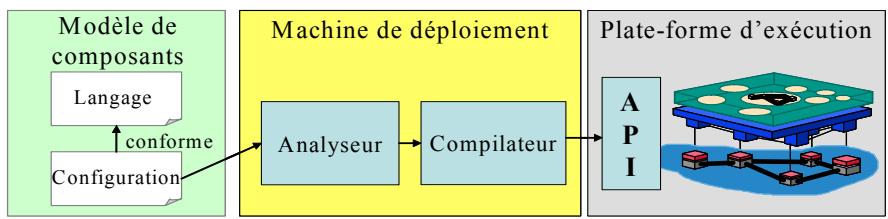

**Figure 1.** *Les éléments fondamentaux d'une plate-forme pour composants logiciels*

En premier lieu, une *configuration applicative* concrète, c'est-à-dire la caractérisation des types de composants, de leurs ports, de la signature des interfaces, des instances de composants à créer, de la configuration de leurs attributs et de la liaison de leurs ports, s'exprime au moyen d'un *langage* spécifique au modèle de composants. Ainsi, dans le modèle de composants CORBA, la description des types de composants, de leurs ports ainsi que la signature des interfaces est exprimée à l'aide du langage *OMG Interface Definition Language* (OMG IDL 3.0) [SCH 04]. Le déploiement, *i.e.* la déclaration des composants à installer et à créer sur chacun des sites concernés et des interconnexions entre leurs ports, est décrit dans le langage XML *Component Assembly Descriptor* (CAD), alors que la configuration des propriétés/attributs des instances est décrite dans le langage XML *Component Property File* (CPF). Dans le modèle de composants Fractal, la signature des interfaces est décrite grâce à un sous ensemble du langage

Java (Fractal IDL) tandis que la caractérisation des types de composants, de leurs ports et du déploiement s'exprime à travers un langage de description d'architecture spécifique : l'ADL Fractal basé sur XML.

En second lieu, une plate-forme pour composants logiciels doit offrir un support d'exécution chargé d'accueillir les composants et des interfaces (ou API) de déploiement. L'API permet le déploiement effectif des composants applicatifs vers ce support d'exécution et est spécifique au modèle.

Enfin, entre ces deux éléments (un langage pour l'expression de configurations concrètes et une plate-forme d'exécution) se trouve la machine de déploiement, généralement constituée d'un analyseur du langage de composants utilisé et d'un compilateur. Le rôle de ce dernier est de traduire les configurations applicatives en appels sur les méthodes offertes par l'API de la plate-forme d'exécution. Par exemple, OpenCCM, notre implantation du modèle de composants CORBA [MAR 01], fournit une infrastructure répartie nommée Distributed Computing Infrastructure (DCI) composée entre autres d'une machine de déploiement implantant l'analyseur/compilateur de descripteurs XML CAD et CPF. C'est la machine de déploiement, propre à chaque plate-forme technologique, que nous souhaitons capitaliser dans notre environnement.

## 3. Notre vision d'un environnement multi personnalités

### 3.1. Motivations

Actuellement, une configuration s'exprime différemment selon la plate-forme à composants utilisée (voir section 2). Ceci présente l'inconvénient majeur de rendre l'architecte dépendant d'une technologie et de son évolution. La migration d'applications d'une plate-forme à une autre n'est pas aisée et nécessite de réécrire entièrement les configurations applicatives dans le langage de la plate-forme cible, ce qui est évidemment coûteux en terme de productivité. Pour permettre aux architectes de capitaliser leurs architectures logicielles, nous proposons donc un environnement multi personnalités dans le but, par exemple, de décrire une configuration avec l'ADL Fractal et déployer l'application sur une plate-forme pour composants CORBA. Ici, l'objectif est de rendre les applications indépendantes des technologies d'expression et d'exécution, ce qui de ce point de vue s'inscrit dans le cadre de la démarche MDA. De plus, la conception et l'implantation d'une machine de déploiement pour un modèle de composants donné reste encore une activité artisanale : il n'existe pas à notre connaissance de méthodologie ni de patrons de conception pour construire de telles machines. Ici, nous proposons une architecture générique pouvant guider les concepteurs de ces machines de déploiement.

*3.2. Architecture*

Pour répondre aux besoins de capitalisation des architectes et des concepteurs, nous avons défini l'architecture d'un environnement rendant la configuration et le déploiement indépendant des choix technologiques. Ceci passe par la capture d'un modèle canonique capable de réifier l'expression d'une configuration et de son déploiement. Ainsi, notre vision d'un tel environnement est illustrée sur la figure 2.

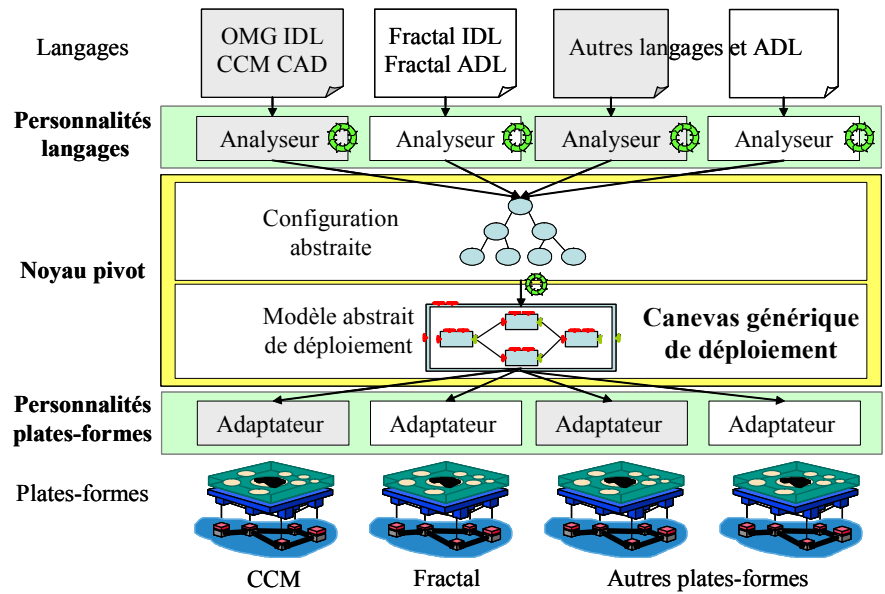

**Figure 2.** *L'architecture de notre environnement multi personnalités*

Cet environnement se compose de trois couches principales :

– La couche *personnalités langages* regroupe les *analyseurs* spécifiques aux langages. Ces *analyseurs* traduisent les configurations concrètes vers une *configuration abstraite* (transformation PSM vers PIM). Cette dernière est un modèle abstrait indépendant de toute technologie composant (*i.e.* un PIM) dont l'objectif est de capturer et réifier en mémoire les configurations applicatives concrètes (types de composants et d'interfaces, configuration des instances, description de leur déploiement), quel que soit le langage d'expression de celles-ci. Cette partie, c'est-à-dire la configuration abstraite et la capture de celle-ci, n'est pas présentée dans cet article et s'inscrit dans nos perspectives de travail.

– Le *noyau pivot* est composé d'un canevas générique qui opérationnalise le processus de déploiement des configurations applicatives, indépendamment des plates-formes cibles, en se basant sur un *modèle abstrait de déploiement* de l'application. Ce modèle de déploiement est obtenu à partir de la *configuration abstraite* via une transformation. Il est décrit plus en détails dans la section 4.

– La couche *personnalités plates-formes* regroupe les *adaptateurs* entre notre *modèle abstrait de déploiement* et les interfaces de déploiement spécifiques aux plates-formes. La section 4.4 décrit les personnalisations pour les plates-formes CORBA et Fractal.

Cette architecture peut être simplifiée lorsque l'objectif est seulement de construire une machine de déploiement pour un modèle de composants spécifique. Dans ce cas, la *configuration abstraite* n'est pas nécessaire et l'*analyseur langage* peut compiler le langage concret directement vers le modèle de déploiement. Cette simplification a déjà été mise en œuvre dans la machine de déploiement pour l'ADL Fractal [FRA 02] où seul notre modèle de déploiement est réutilisé.

### 3.3. Bénéfices

Les bénéfices de notre approche sont nombreux. Du point de vue des architectes, elle offre la possibilité de décrire une configuration dans divers langages et de la projeter vers différentes plates-formes. La migration de configurations entre modèles de composants peut être envisagée en ajoutant des transformations traduisant des configurations abstraites vers les langages concrets. D'un point de vue ingénierie, notre approche propose un canevas logiciel concret et réutilisable pour construire des machines de déploiement (voir section 4) et elle permet de clairement séparer les préoccupations liées au traitement des langages, de l'exécution du déploiement et de son opérationnalisation sur des plates-formes concrètes. D'un point de vue purement scientifique, notre approche vise à capturer le modèle abstrait de la configuration et du déploiement indépendamment de toute technologie. A terme, cela nous donnera la possibilité de comparer l'expressivité des langages de composants et leurs limitations, puis d'identifier les manques dans les interfaces de déploiement d'une plate-forme spécifique et éventuellement d'identifier ainsi un modèle canonique « idéal » des composants logiciels. Enfin, la réification explicite des configurations applicatives offre un support approprié pour s'attaquer à la reconfiguration des applications durant l'exécution.

## 4. Le modèle de déploiement multi plates-formes de composants

### 4.1. Le déploiement sous la forme de graphes de tâches

Le déploiement automatique d'une application à base de composants logiciels consiste à exécuter un ensemble de tâches élémentaires comme le téléchargement des binaires de composants sur leur site d'exécution, le chargement de ces binaires dans la mémoire des serveurs d'applications, l'instanciation des binaires pour créer des instances de composants, la configuration des attributs métiers des instances, l'établissement des liaisons entre les interfaces requises et offertes des différents composants, puis le démarrage des instances de composants. L'ensemble de ces tâches doit être coordonné ou orchestré dans un ordre précis. Par exemple, la configuration d'un attribut d'une instance de composant ne doit avoir lieu que lorsque la création de l'instance a été effectuée, cette dernière ne pouvant avoir lieu que lorsque le téléchargement du binaire du composant et son chargement en mémoire ont été réalisés au préalable.

Ainsi, notre approche repose sur la modélisation du déploiement sous la forme d'un graphe de tâches élémentaires (*i.e.* un workflow) orchestrées automatiquement en fonction des dépendances entre ces tâches. Ce modèle définit différents types de tâches représentant les actes élémentaires et strictement nécessaires au déploiement d'une application à base de composants logiciels, à savoir l'installation d'une fabrique de composant, l'instanciation d'un composant, la configuration d'un attribut, l'obtention d'une interface offerte par un composant, la liaison avec une interface requise, l'ajout d'un sous composant à un composant et enfin le démarrage effectif d'un composant. Les différents types de tâches sont modélisés sous la forme de composants logiciels (voir section 4.3). Ici le paradigme composant nous permet d'identifier les propriétés de configuration de chaque tâche et leurs dépendances. Les dépendances sont modélisées par des liaisons entre les interfaces offertes et requises des composants représentant les tâches.

*4.2. Le canevas générique d'orchestration de tâches*

Afin d'ordonnancer automatiquement et correctement l'exécution des tâches de déploiement, nous avons conçu un canevas générique d'orchestration de tâches. Celles-ci sont modélisées sous la forme d'un composant Fractal devant offrir l'interface ***Task***. Chaque composant doit implanter l'opération ***execute()*** en fonction de la nature de la tâche à accomplir. Les composants doivent être placés dans un composite d'orchestration offrant lui-même l'interface ***Task***. Ainsi, un composant orchestration peut en contenir récursivement d'autres. Le composite ***Orchestration*** doit être vu comme un opérateur particulier de composition de composants ***Task***. Lorsque la méthode ***execute()*** d'un composite ***Orchestration*** est invoquée celui-ci va exécuter les tâches dans un ordre respectant leurs dépendances, celles-ci étant exprimées sous la forme de liaisons entre les composants. La figure 3 illustre le graphe des tâches nécessaires au déploiement d'une application simple faisant intervenir deux composants (un composant *Serveur* possède une interface fournie qui peut être utilisé par un composant *Client*, par l'intermédiaire de son interface requise), c'est-à-dire l'assemblage des composants de déploiement utilisés pour déployer l'application. Ici, l'orchestration exécute les tâches d'installation des binaires des composants *Serveur* et *Client*, puis les tâches d'instanciation, de configuration des attributs *nom*, de liaison des interfaces et enfin d'initialisation finale. La liaison entre les deux composants se déroule en deux temps : l'obtention de l'interface offerte par le composant *Serveur* puis la liaison avec l'interface requise par le composant *Client*.

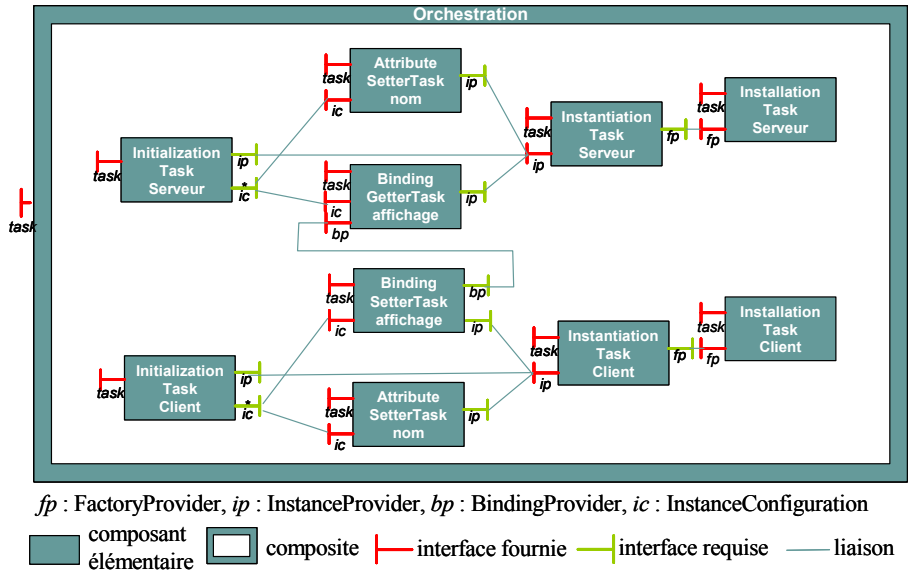

**Figure 3.** *Le graphe des tâches de déploiement pour l'application Client/Serveur*

Comme on peut le constater sur la figure 3, il n'existe pas un unique ordonnancement des tâches à exécuter. Certaines peuvent être exécutées en parallèle ou sérialisées, comme par exemple les installations, les instanciations et les initialisations, sans pour autant changer le résultat final de l'exécution du graphe de tâches du déploiement.

Techniquement, la méthode ***execute()*** du composite ***Orchestration*** crée une structure de données contenant pour chaque tâche la liste *TP* des tâches dont elle dépend et la liste *TS* des tâches qui en dépendent. Ces listes sont construites par utilisation des opérations d'introspection structurelle offertes par le modèle de composants Fractal (*i.e.* les interfaces de contrôle *Component*, *ContentController* et *BindingController*). De manière itérative, les tâches dont la liste *TP* est vide sont alors exécutées et leur référence est retirée de la liste *TP* de toutes les tâches de leur liste *TS*. Ainsi elles sont exécutées selon un ordonnancement correct vis-à-vis de leurs dépendances et les cycles peuvent être détectés simplement lorsqu'il reste des tâches à exécuter à la fin de l'algorithme. Un prototype de ce canevas générique d'orchestration de tâches est disponible en logiciel libre sous la licence GNU LGPL à l'adresse http://forge.objectweb.org/projects/deployment.

*4.3. Les composants de déploiement*

Notre expérience sur les modèles de composants Fractal et CORBA nous a amené à identifier un ensemble minimaliste de tâches élémentaires et strictement nécessaires pour le déploiement de configurations de composants logiciels. Cet ensemble est constitué de sept types de composants offrant quatre types d'interfaces. La liste suivante identifie les différentes interfaces offertes ou requises par les différents types de composants de déploiement :

– ***FactoryProvider*** : cette interface fournit la référence d'une fabrique de composants (opération ***getFactory***).

– ***InstanceProvider*** : cette interface fournit la référence d'une instance de composant (opération ***getInstance***).

– ***BindingProvider*** : cette interface fournit la référence d'une interface offerte par une instance de composant (opération ***getBinding***).

– ***InstanceConfiguration*** : cette interface vide permet de taguer les tâches dédiées à la configuration d'une instance de composant.

La liste suivante identifie les différents types de composants de déploiement en précisant pour chacun son rôle, ses propriétés, ses interfaces requises et offertes :

– ***InstallationTask*** : ce type modélise l'installation d'une fabrique de composants, c'est-à-dire la mise en place d'un quelconque moyen permettant de créer des instances de composants et offre l'interface ***FactoryProvider***.

– ***InstantiationTask*** : ce type modélise la création d'une instance de composants, requiert l'interface ***FactoryProvider*** et offre l'interface ***InstanceProvider***.

– ***AttributeSetterTask*** : ce type modélise la configuration d'un attribut, possède les propriétés *Name* et *Value* (nom et valeur de l'attribut), requiert l'interface ***InstanceProvider*** et, de même que pour les trois types suivants, offre l'interface ***InstanceConfiguration***.

– ***BindingGetterTask*** : ce type modélise l'obtention d'une interface offerte, possède la propriété *Name* (nom de l'interface offerte), requiert l'interface ***InstanceProvider*** et offre l'interface ***BindingProvider***.

– ***BindingSetterTask*** : ce type modélise l'établissement de la liaison d'une interface requise (avec une interface offerte), possède la propriété *Name* et requiert une interface ***InstanceProvider*** et une interface ***BindingProvider***.

– ***AddComponentTask*** : ce type, nécessaire uniquement pour les modèles de composants hiérarchiques comme Fractal, modélise l'ajout d'un sous-composant. Il possède une propriété *Name* (identifiant du sous-composant) et requiert deux interfaces ***InstanceProvider*** permettant d'obtenir les références des deux instances de composant (le contenant et le contenu).

– ***InitializationTask*** : ce dernier type modélise la fin de la configuration des fonctionnalités d'une instance et permet d'initier le démarrage effectif de l'instance de composant, requiert une interface *InstanceProvider* et un ensemble de références vers les interfaces *InstanceConfiguration* devant être exécutées avant.

### 4.4. La personnalisation pour les plates-formes de composants

La personnalisation de notre modèle pour une plate-forme spécifique nécessite de spécialiser les composants de déploiement en fonction des API de déploiement et consiste à implanter la méthode *execute()* des différents composants. Le tableau 1 illustre les spécialisations pour les modèles Fractal et CORBA.

| Type de composants | CCM | Fractal |
|---|---|---|
| **InstallationTask** | Opération *install_home* de l'interface *Container* obtenue via les interfaces *ComponentServer* et *ServerActivator* | Opération *newFcInstance* de l'interface *GenericFactory* du composant d'amorçage |
| **InstantiationTask** | Opération *create_component* de l'interface *CCMHome* | Opération *newFcInstance* de l'interface *Factory* |
| **AttributeSetterTask** | Opération d'affectation de l'attribut sur l'interface du composant | Opération d'affectation de l'attribut sur l'interface *AttributeController* |
| **BindingGetterTask** | Opérations *provide_facet* et *get_consumer* de l'interface *CCMObject* | Opération *getFcInterface* de l'interface *Component* |
| **BindingSetterTask** | Opérations *connect* et *subscribe* de l'interface *CCMObject* | Opération *bindFc* de l'interface *BindingController* |
| **AddComponentTask** | Non valable dans ce modèle | Opération *addFcSubComponent* de l'interface *ContentController* |
| **InitializationTask** | Opération *configuration_complete* de l'interface *CCMObject* | Opération *startFc* de l'interface *LifeCycleController* |

**Tableau 1.** *La spécialisation des composants de déploiement pour CCM et Fractal*

Ce tableau montre quelles sont les opérations sur les interfaces de l'API de déploiement à invoquer afin de réaliser les tâches des composants de déploiement. Le modèle de composants CORBA spécifie dans le module *Components* un certain nombre d'interfaces pour réaliser les opérations du processus de déploiement : installation des binaires, création des fabriques, instanciation des composants, configuration de leurs attributs et interconnexion de leurs ports. Dans le modèle Fractal, chaque composant peut offrir un ensemble d'interfaces de contrôle (*Component*, *LifeCycleController*, *AttributeController*, *BindingController* et *ContentController*) permettant de configurer ses attributs, de lier ses interfaces, de construire des composants composites et de le démarrer. De plus, un composant d'amorçage (*Bootstrap*) permet de créer des fabriques de composants.

Ainsi, pour l'implantation de la spécialisation du composant ***InitializationTask*** par exemple, il s'agit d'obtenir la référence du composant, de type *CCMObject* pour le modèle de composant CORBA et *Component* pour Fractal, puis d'invoquer l'opération correspondante pour le démarrer, *i.e.* respectivement *configuration_complete()* et *startFc()*.

Finalement, notre modèle de déploiement est facilement personnalisable pour des plates-formes de composants sophistiquées telles que CORBA et Fractal. La personnalisation des tâches pour le modèle Fractal a d'ores et déjà été écrite par Eric Bruneton (France Télécom R&D) dans l'implantation de l'ADL Fractal et est disponible en logiciel libre [FRA 02]. La personnalisation pour le modèle de composants CORBA est en cours de prototypage et devrait à terme être incluse dans notre plate-forme logicielle libre OpenCCM [MAR 01, OPE 02].

## 5. Conclusion et perspectives

Dans cet article, nous avons soulevé le problème de la capitalisation par les architectes de l'expression de la configuration et du déploiement des applications à base de composants logiciels. En effet, en dehors du cadre d'une technologie, cette capitalisation n'existe pas. Notre approche pour répondre à ce besoin consiste à proposer un environnement multi personnalités permettant aux architectes d'exprimer la configuration et le déploiement d'une application à base de composants indépendamment d'un choix de langage et de plate-forme. L'intérêt de notre approche réside dans le fait qu'elle soustrait les architectes au problème de la dépendance vis-à-vis d'un modèle de composants (donc de ses limites et de son évolution, voire de sa disparition) et permet une meilleure réutilisation des applications métiers. De plus, elle n'ajoute pas de contraintes comme l'apprentissage d'un nouveau langage ou modèle abstrait dans la mesure où les architectes peuvent continuer à utiliser leur modèle favori, mais au contraire, simplifie les migrations inter plates-formes. La contribution de cet article a été de décrire l'architecture d'un tel environnement, le modèle de déploiement sous-jacent et sa personnalisation pour les modèles de composants CORBA et Fractal.

Les perspectives pour ce travail vont dans plusieurs directions. Nous souhaitons maintenant nous attaquer à la définition précise du modèle abstrait de configuration de notre environnement. Pour cela, nous comptons fusionner plusieurs sources d'inspiration, à savoir principalement le modèle abstrait Fractal et le récent modèle de déploiement et de configuration de l'OMG. Parallèlement, un travail conséquent doit être accompli pour spécifier et implanter différentes personnalisations langages et plates-formes de notre environnement. Ainsi, l'ajout de nouveaux types de composants de déploiement permettra de supporter des configurations applicatives basées sur des concepts de déploiement différents. Un de nos objectifs serait d'appliquer notre approche à d'autres technologies comme les EJB et les Web Services. Enfin, nous souhaitons améliorer notre canevas générique d'orchestration afin d'offrir la possibilité d'annuler une ou plusieurs tâches et de revenir à un état précédent. Ceci peut en effet s'avérer fort utile en cas de panne survenant durant le processus de déploiement d'une application.

**6. Bibliographie**